\documentclass[twoside]{article}
\usepackage{multicol}
\usepackage[jce]{iwce}     
\usepackage[intlimits]{amsmath}
\usepackage{amssymb, amsmath}

\usepackage{graphicx}
\usepackage{overpic}
\usepackage{color}

\begin{document}

\title{Transport in Silicon Nanowires: Role of Radial Dopant Profile}

\IWCEauthorsFirst{TROELS MARKUSSEN$^1$,\;\;RICCARDO RURALI$^{2}$,\;\;ANTTI-PEKKA JAUHO$^{1,3}$,\;\;MADS
BRANDBYGE$^1$} 
\setHeadings{Markussen}{Transport in Silicon Nanowires: Role of Radial Dopant Profile}
\IWCEaddressFirst{$^1$MIC - Department of Micro and Nanotechnology, NanoDTU, Technical University of Denmark\\
$^2$Departament d'Enginyeria Electronica, Universitat Autonoma de Barcelona, Spain\\
$^3$Laboratory of Physics, Helsinki University of Technology, P.O. Box 1100 FI-02015 HUT,
Finland } 
\email{troels.markussen@mic.dtu.dk,\;\;Riccardo.Rurali@uab.cat,\;\;antti@mic.dtu.dk,\;\;mbr@mic.dtu.dk}
\preparetitle

\begin{IWCEabstract}
We consider the electronic transport properties of phosphorus (P) doped silicon nanowires (SiNWs).
By combining {\it ab initio} density functional theory (DFT) calculations with a recursive
Green's function method, we calculate the conductance distribution of up to 200 nm
long SiNWs with different distributions of P dopant impurities.
We find that the radial distribution of the dopants influences the conductance
properties significantly: Surface doped wires have longer mean-free paths and smaller
sample-to-sample fluctuations in the cross-over from ballistic to diffusive transport.
These findings can be quantitatively predicted in terms of the scattering properties of the single dopant atoms, implying that relatively simple calculations are sufficient in practical device modeling
\end{IWCEabstract}

\begin{IWCEkeywords}
Silicon nanowires, dopant scattering, DFT transport calculations, mean-free path,
sample-to-sample fluctuations
\end{IWCEkeywords}

\begin{multicols}{2}
\intro{Introduction}

The continuous reductions in feature sizes in the semiconductor
industry have generated much interest in novel one-dimensional
nanostructures such as silicon nanowires (SiNWs)
\cite{Lieber,Patolsky}. Scattering by
defects or dopants will be increasingly important with decreasing
sizes, and at the same time sample-to-sample variations become a
crucial issue: when the device length and the mean free path are
comparable, and shorter than the coherence length, variations of the
positions of the individual dopant atoms can affect the conductance
of the wire significantly. Accurate models for the electronic
transport properties of nanowires including sample-variations are
clearly desirable.

The mathematical theory of the conductance of disordered quasi one-dimensional systems has
reached a high level of understanding in the diffusive as well as the localization regime
\cite{LeeRamakrishnan,Beenakker}. The diffusive regime is characterized by $l_e<L<\xi$
with $L,\,l_e$ and $\xi$ being the sample length, mean free path (MFP) and localization
length, while in the localization regime $L>\xi$. The universal conductance fluctuations
(UCF) is a remarkable property of disordered systems: In the diffusive regime the
sample-to-sample variations are described by a Gaussian distribution with universal width
of the order of $e^2/h$ independent on the shape of the conductor and type of disorder
\cite{LeeStonePRL1985}.

Several recent theoretical works used density-functional theory (DFT) to consider the
energetics of dopants in SiNWs \cite{Blase_PRL_2005,Peelears}. Further, Fernandez-Serra et
al.\cite{Blase_NanoLett_2006} considered scattering properties of {\it single} phosphorus
(P) dopants. In a recent work \cite{MarkussenPRL2007} we studied sample-averaged
conductance properties ($\langle G\rangle$, std($G$), MFP) in {\it long} SiNWs containing
a random distribution of either B or P dopants along the wire. In this work we continue
the analysis focusing more on the conductance {\it distribution} and on the influence of
the radial dopant profile.


\section{Method} \label{Method}
\noindent
In this work we model the conductance properties of SiNWs with randomly distributed P dopant atoms. The wires are oriented along the $[1\,0\,0]$ direction and are surface passivated by hydrogen atoms to avoid dangling bonds. The Hamiltonians describing the pristine wire as well as regions containing dopant impurities are obtained using {\it ab initio} density-functional theory (DFT) calculations \cite{Siesta}. The DFT calculations on the doped regions use supercells containing nine wire unit cells (837 atoms) with a total length of 50.4 \AA~and with the single dopant atom placed in the middlemost unit cell. The atomic positions of all the atoms in the central unit cell containing the dopant atom as well as the two neighboring unit cells have been fully relaxed, until the maximum force was smaller than 0.04 eV/\AA. The large super cell ensures that the there is no dopant-dopant interaction and that the electronic structure at the supercell-supercell interface is very close to that of the pristine wire. The length and energy dependent conductance of a long wire with randomly placed P atoms is calculated using a standard recursive Green's function (GF) method \cite{MarkussenPRB2006}. In each recursion step, we add a unit cell from either the pristine wire or from a region around a dopant atom. At each energy we do this calculation for 560 different realizations of the dopant positions to obtain the conductance distribution. By changing the relative occurrence of  different dopant positions, we can model different radial dopant profiles. In all calculations we use an average dopant-dopant separation $d=10\,$nm corresponding to a bulk doping density of $n\approx 10^{19}$~cm$^{-3}$. Since P dopants are n-type we consider only energies in the conduction band and measures all energies relative to the conduction band edge, $E_c$.

\section{Single-dopant results and estimates} \label{Singles}
\noindent
\Fig{singleTrans} shows the transmission vs. energy through
infinite wires containing only a single P dopant atom placed at five
different substitutional positions (1-5) indicated in the inset. The P dopants located in the bulk positions (1-3) generally scatter the electrons more than the surface positions (4-5).

In a recent work \cite{MarkussenPRL2007} we showed that the sample-averaged properties, $\langle G\rangle$, std($G$) and MFP could be accurately predicted using information only from the single-dopant transmissions shown in \Fig{singleTrans}. We define an average scattering resistance
\begin{equation}
\label{R_s}
\langle R_s(E)\rangle=\sum_{i=1}^M\frac{p_i}{G_i(E)}-\frac{1}{G_0(E)},
\end{equation}
where $\sum_{i=1}^Mp_i=1$ and $1/G_0(E)= h/(N(E)\,2\,e^2)=R_c(E)$ is the contact
resistance with $N(E)$ being the number of conducting channels at energy $E$. The first
term on the right is the average of the single-dopant conductances taking the probability,
$p_i$, of each dopant position into account. In the present study $i=1\hdots 5$,
corresponding to the five different dopant positions shown in the inset in
\Fig{singleTrans}. Denoting the average dopant-dopant distance, $d$, the average
resistance of a wire of length, $L$, is estimated as
\begin{equation}
\label{R_single}
R(E,L) = R_c(E) + \langle R_s(E)\rangle\,L/d.
\end{equation}
\begin{IWCEfigure}
\centering\begin{overpic}[width=.85\linewidth]{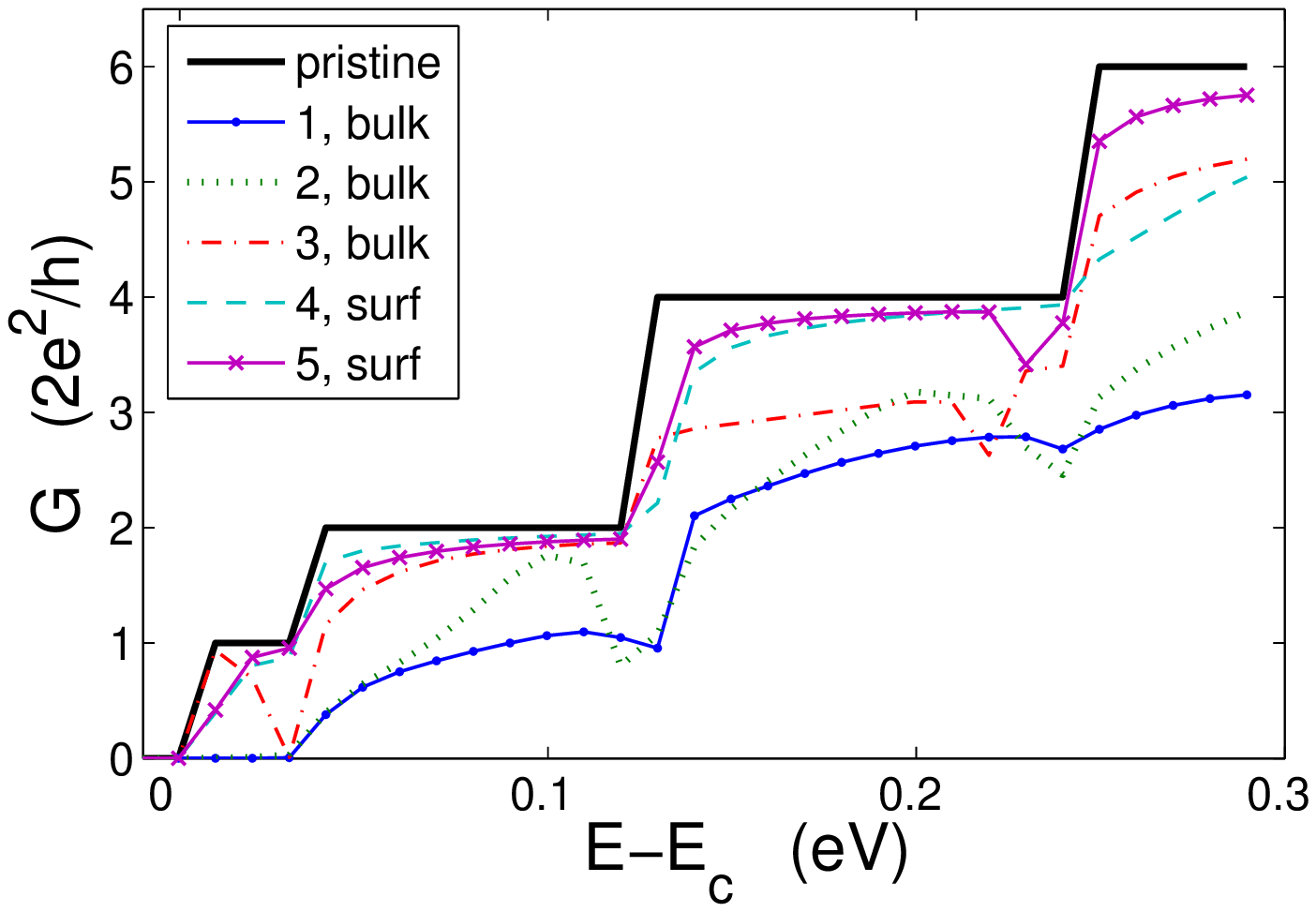}
    \put(67.,12){\includegraphics[width=.18\linewidth]{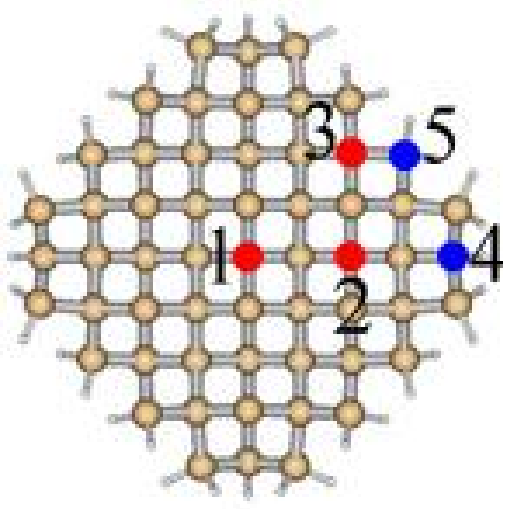}}
\end{overpic}
\caption{\label{singleTrans} Single dopant conductance vs. energy. Each curve correspond to a specific position of the dopant atom, as indicated in the inset showing a cross section of the wire with the possible dopant positions marked.}
\end{IWCEfigure}
From the linear region of the $\langle R\rangle$ vs. $L$ curves (see  \Fig{R_and_std} (a))
one can extract the energy dependent MFP, $l_e(E)$ through the relation
\begin{equation}
\label{MFP_eq}
\langle R(E,L)\rangle = R_c(E) + R_c(E)\,L/l_e(E).
\end{equation}
Comparing (\ref{R_single}) with (\ref{MFP_eq}) we obtain an expression for the MFP given only by the single-dopant transmissions, the doping-dopant distance, and the dopant distribution:
\begin{equation}
\label{MFP_singles}
\tilde{l}_e(E) = d\,\frac{R_c(E)}{\langle R_s(E)\rangle}.
\end{equation}
We emphasize that the relations above only hold for wire lengths shorter than the
localization length, $\xi$. We can calculate $\xi$ using the relation \cite{Beenakker},
\begin{equation}
\label{loc_length} \xi(E)= \frac{1}{2}(N(E)+1)l_e(E)\,,
\end{equation}
which now can be estimated using the single-dopant result, $\tilde{l}_e(E)$
\cite{MarkussenPRL2007}.

\section{Long wire results} \label{Long-wire}
\noindent
\Fig{hist} shows the conductance distribution for different wire lengths
at energy $E=0.16\,$eV (a)-(d) and $E=0.01\,$eV (e)-(h).
The radial dopant distribution is uniform, i.e. the positions 1-5 are equally likely.
The conductance distribution at the higher energy (left) develops from a quasi-ballistic regime (a) with a few peaks in the histogram, to the diffusive regime (b)+(c) and finally to the beginning of the localization regime (d).

In the diffusive regime the conductance distribution is well described by a Gaussian with the standard deviation being close to the universal conductance fluctuation (UCF) value of $0.37\cdot\frac{2e^2}{h}$. In the ballistic and diffusive regimes, the sample averaged resistance $\langle R\rangle=1/\langle G\rangle$ increases linearly with wire length, as illustrated in \Fig{R_and_std} (a) (circles). For wire length $L>150\,$nm the resistance starts to increase above the initial linear behavior, thus entering the localization regime. Notice that the localization sets in when $\langle R\rangle \approx 1 \cdot\frac{h}{2e^2}$ in agreement with Thouless' argumentation \cite{Thouless}. 
The resistance of the surface-doped wires \Fig{R_and_std} (a) (stars) increases linearly throughout the length range and stays in diffusive regime. The two solid lines in \Fig{R_and_std} (a) obtained using (\ref{R_single}) clearly resemble the sample-averaged values (circles and stars) in the linear regimes. This demonstrates, that the average resistance in the ballistic- and diffusive regimes can be accurately predicted for various dopant distributions only based on the single-dopant scattering properties.
\begin{IWCEfigure}
\centering\includegraphics[width=0.95\textwidth]{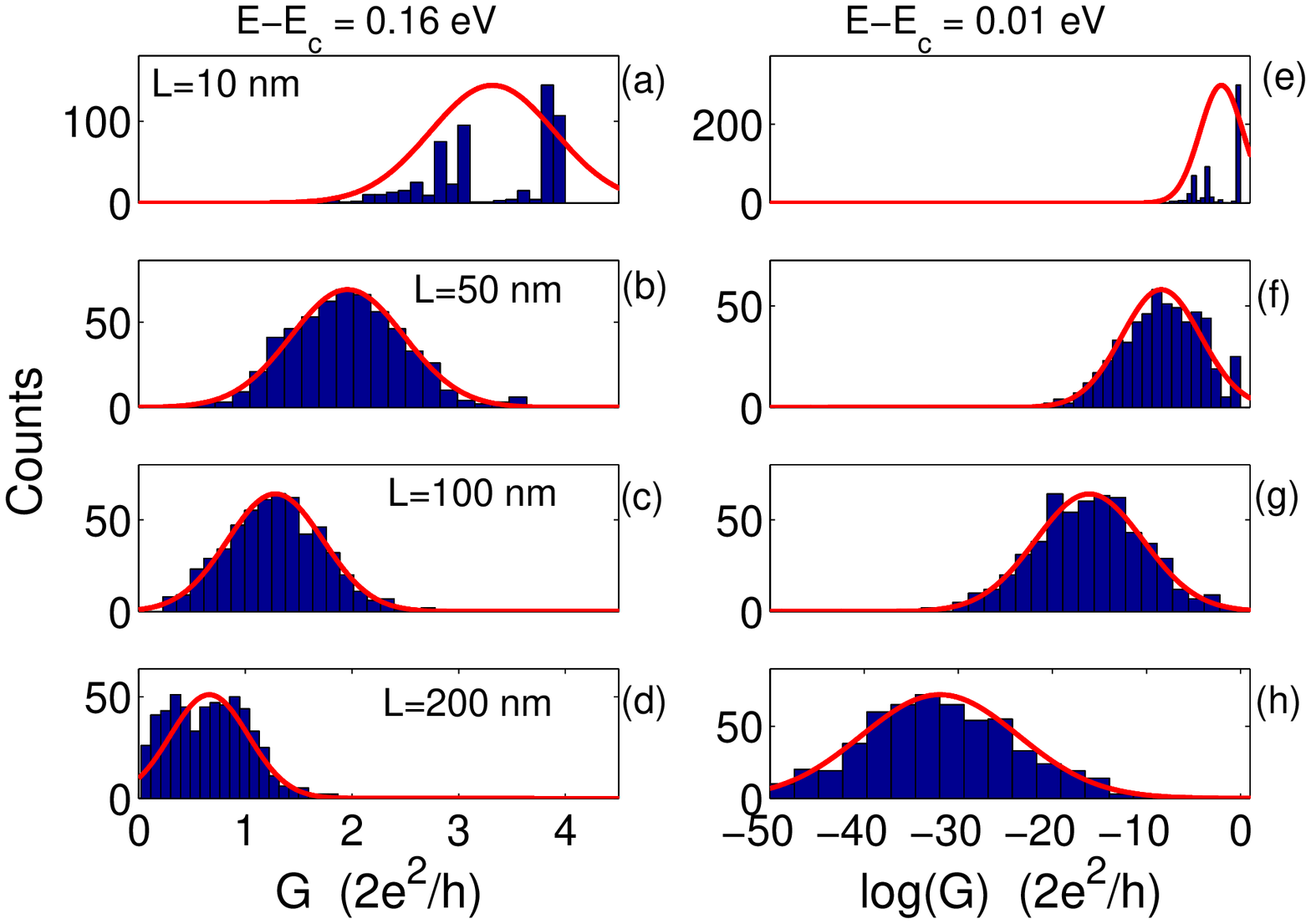}
\caption{\label{hist} Conductance distributions at different wire lengths (rows) for a uniform radial dopant distribution  at energy $E=0.16\,$eV (left) and $E=0.01\,$eV (right).}
\end{IWCEfigure}
In the distribution plots the diffusive-localization transition starts when the low
conductance tails of the histograms come close to zero and starts to 'pile up'. Eventually
the distribution develops into a log-normal distribution. This is further illustrated in
the right part (e)-(h) showing the distribution of log($G$) at the energy $E=0.01\,$eV
just above the conduction band edge.
\begin{IWCEfigure}
\centering\includegraphics[width=.9\linewidth]{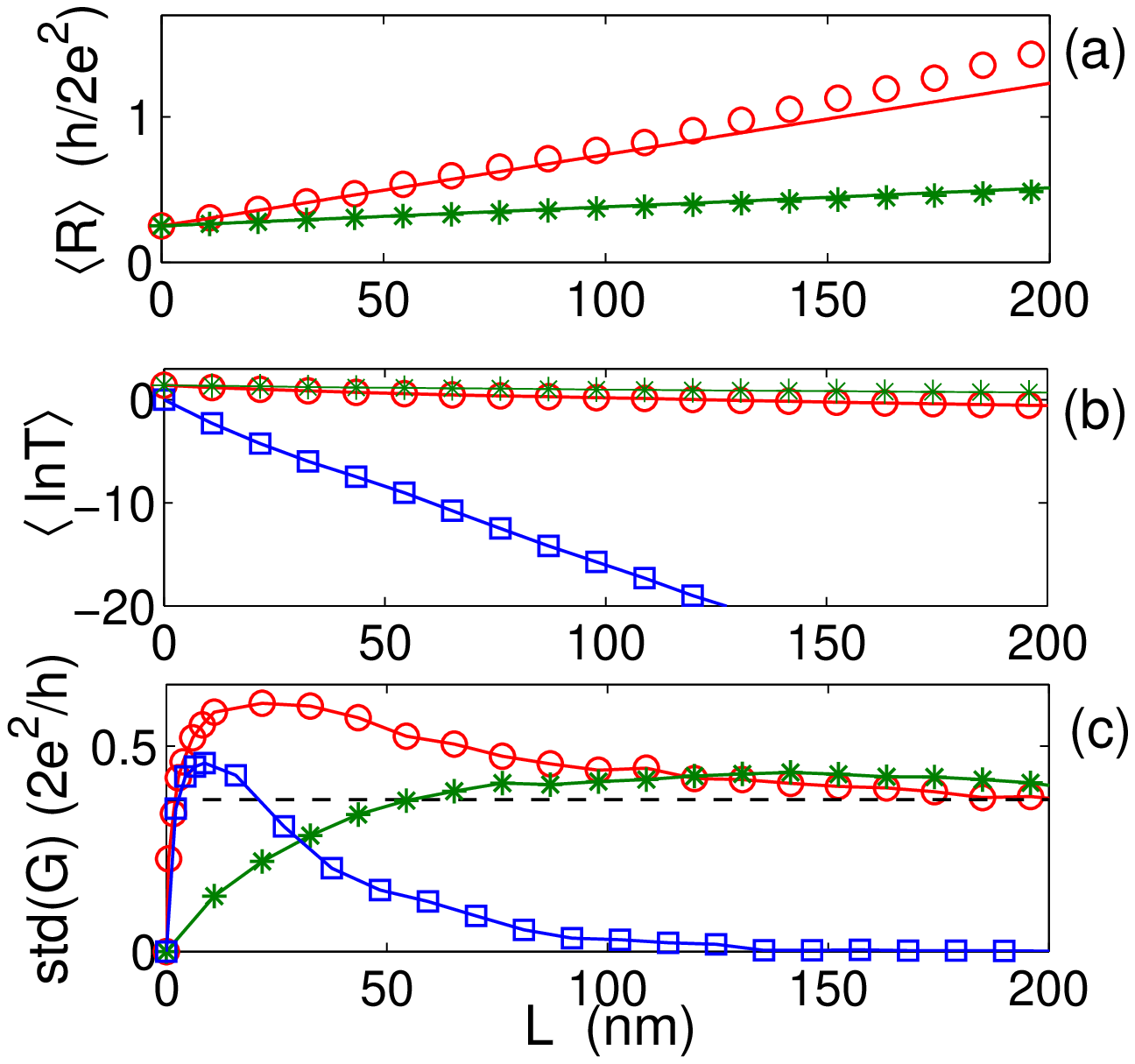}
\caption{\label{R_and_std} (a): Average resistance at $E=0.16\,$eV of uniformly doped wires (circles) and surface doped wires (stars). The solid lines are estimates based on the single dopant transmissions (\Fig{singleTrans}) using \eqref{R_single}. The MFP is found from the linear regions giving  $l_e=47\,$nm for the uniform distribution and  $l_e=190\,$nm for the surface doped wires. (b): ln($T$) vs. $L$ for the two different distributions in (a) (circles and stars) and for uniform distribution at $E=0.01\,$eV (squares). (c): std($G$) vs. $L$. Same symbols as above. }
\end{IWCEfigure}
The transmission, $T=G\cdot\frac{h}{2e^2}$, decreases as $T(L)\sim\exp(-L/\xi)$ for wires in the localization regime. This is illustrated in \Fig{R_and_std} (b) showing $\langle \ln T \rangle$ vs. $L$ at $E=0.01\,$eV (squares) and at $E=0.16\,$eV for the uniformly doped wires (circles) and the surface doped wires (stars).  The localization length is obtained from the slopes in the final linear regions. At $E=0.01\,$eV we get $\xi=7\,$nm and for the uniformly doped wire at $E=0.16\,$eV we find $\xi=133\,$nm. The surface doped wire does not reach a linear region in the considered length range and we can only say that $\xi>200\,$nm. The calculated localization lengths generally agree with the relation (\ref{loc_length}) giving $\xi=8$ and $123\,$nm at $E=0.01$ and $0.16\,$eV respectively.

The length dependences of the sample-to-sample variations, std($G$), shown in \Fig{R_and_std} (c) are qualitatively different for the two radial dopant distributions. The purely surface doped wires (stars) approach the UCF level (horizontally dashed line) from below. On the other hand, the uniformly doped wires (circles) reaches a clear maximum before the UCF level is approached from above. A similar maximum is observed at $E=0.01\,$eV (squares). Both maxima occur around $L\approx l_e/2$, which seems to be a quite general result independent of energy and dopant concentration \cite{MarkussenPRL2007}.

\begin{IWCEfigure}
\centering\includegraphics[width=.95\linewidth]{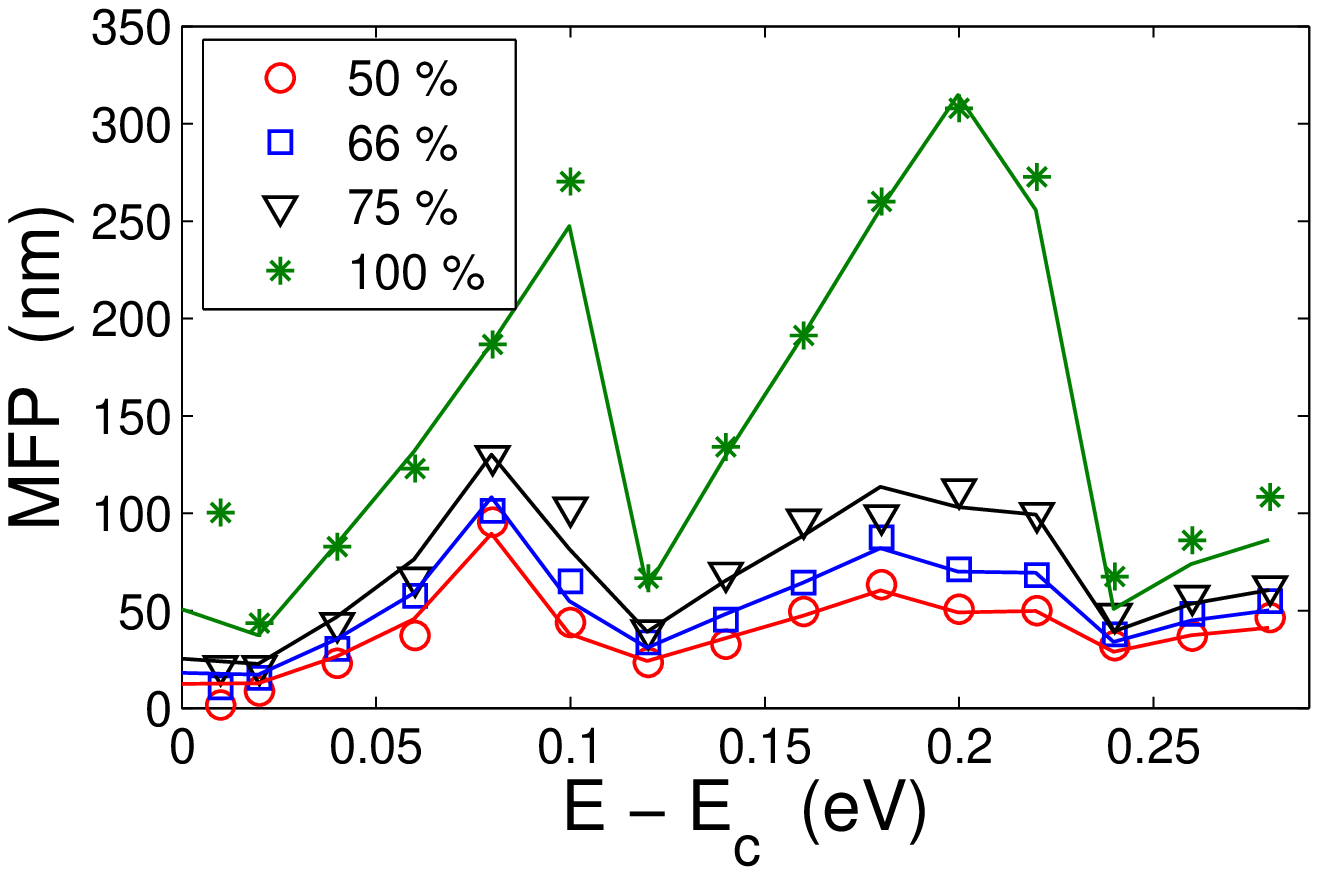}
\caption{\label{mfp_fig} MFP vs. energy for different radial dopant profiles. The numbers in the legend refers to the fraction of the dopant atoms located at the surface. All solid lines are results obtained from the single-dopant transmissions.}
\end{IWCEfigure}
Using (\ref{MFP_eq}) we extract the MFP from the linear regions in \Fig{R_and_std} (a). By doing this at many energies, we obtain the energy dependent MFP shown in \Fig{mfp_fig}. The different markers correspond to different surface-dopant fractions. The solid lines are obtained from the single-dopant calculations using (\ref{MFP_singles}) and correspond to the same surface dopant fraction as the closely lying markers. We see that the single-dopant estimates accurately resemble the sample-averaged results over the entire energy range.
At all energies, the MFP increases with increasing surface dopant fraction, and at pure surface doping the MFP is on average more than three times larger than for uniform doping. The large increase in MFP for surface doped wires might be useful for device applications. However, there are potential problems with dopants close to the surface as they can be passivated by extra hydrogen atoms and thereby no longer provide extra carriers \cite{Blase_NanoLett_2006}.

\section{Conclusion}
\noindent
By combining ideas of scaling theory and universal conductance fluctuations with DFT and Landauer formalism we have analyzed the conductance properties of SiNWs. We have studied the transport in phosphorus doped SiNWs by computing the sample averaged conductance, mean free path (MFP), localization length, and sample-to-sample variations as a function of Fermi energy, radial doping distribution, and wire length. The MFP was significantly increased in surface doped wires as compared to more uniform dopant distributions. Moreover, the sample-to-sample variations at wire lengths $L<l_e$ were smaller in the surface doped wires.
These results can be understood and accurately estimated from the scattering properties of the {\it single} dopants, implying that relatively simple calculations are sufficient in practical nanoscale device modeling.

%
%
%
%
%
%

\section*{Acknowledgment}
\noindent We thank the Danish Center for Scientific Computing (DCSC) for providing
computer resources. APJ is grateful to the FiDiPro program of the Finnish Academy for
support during the final stages of this work. RR acknowledges financial support from
Spain's Ministerio de Educaci\'{o}n y Ciencia Juan de la Cierva program and funding under
Contract No. TEC2006-13731-C02-01.

\bibliographystyle{IEEEbib_jce}

\end{multicols}
\end{document}